\documentclass{article}

\PassOptionsToPackage{numbers, compress}{natbib}

\usepackage[final]{neurips_2024}

\usepackage{graphicx}
\usepackage[utf8]{inputenc} 
\usepackage[T1]{fontenc}    
\usepackage{hyperref}       
\usepackage{url}            
\usepackage{booktabs}       
\usepackage{amsfonts}       
\usepackage{nicefrac}       
\usepackage{microtype}      
\usepackage{xcolor}         
\usepackage{textcomp}       
\usepackage{natbib}         

\usepackage{listings}
\usepackage{todonotes}

\usepackage{framed}
\usepackage{xcolor}

\usepackage{caption}
\usepackage{longtable}    

\title{Automated, LLM enabled extraction of synthesis details for reticular materials from scientific literature}

\author{
  Viviane Torres da Silva\\
  IBM Research\\
  \texttt{vivianet@br.ibm.com}\\
  \And
  Alexandre Rademaker \\
  IBM Research \\
  \texttt{alexrad@br.ibm.com} \\
  \And
  Krystelle Lionti\\
  IBM Research \\
  \texttt{klionti@us.ibm.com} \\
  \And
  Ronaldo Giro\\
  IBM Research \\
  \texttt{rgiro@br.ibm.com} \\
  \And
  Geisa Lima\\
  IBM Research \\
  \texttt{Geisa.Lima@ibm.com} \\
  \And
  Sandro Fiorini\\
  IBM Research \\
  \texttt{srfiorini@ibm.com} \\
  \And
  Marcelo Archanjo\\
  IBM Research \\
  \texttt{marcelo.archanjo@ibm.com} \\
  \And
  Breno W. Carvalho\\
  IBM Research \\
  \texttt{brenow@ibm.com} \\
  \And
  Rodrigo Neumann\\
  IBM Research \\
  \texttt{rneumann@br.ibm.com} \\
  \And
  Anaximandro Souza\\
  IBM Research \\
  \texttt{anaximandrosouza@ibm.com} \\
  \And
   João Pedro Souza\\
  Idiap Research Institute \\
  \texttt{joao.gandarela@idiap.ch} \\
  \And
  Gabriela de Valnisio\\
  IBM Research \\
  \texttt{gvalnisio@ibm.com} \\
  \And
  Carmen Nilda Paz\\
  IBM Research \\
  \texttt{cpaz@br.ibm.com} \\
  \And
  Renato Cerqueira\\
  IBM Research \\
  \texttt{rcerq@br.ibm.com} \\
  \And
  Mathias Steiner\\
  IBM Research \\
  \texttt{mathiast@br.ibm.com} \\
}  

\begin{document}

\maketitle

\begin{abstract}

  Automated knowledge extraction from scientific literature can potentially accelerate materials discovery. We have investigated an approach for extracting synthesis protocols for reticular materials from scientific literature using large language models (LLMs). To that end, we introduce a Knowledge Extraction Pipeline (KEP) that automatizes LLM-assisted paragraph classification and information extraction. By applying prompt engineering with in-context learning (ICL) to a set of open-source LLMs, we demonstrate that LLMs can retrieve chemical information from PDF documents, without the need for fine-tuning or training and at a reduced risk of hallucination. By comparing the performance of five open-source families of LLMs in both paragraph classification and information extraction tasks, we observe excellent model performance even if only few example paragraphs are included in the ICL prompts. The results show the potential of the KEP approach for reducing human annotations and data curation efforts in automated scientific knowledge extraction.
\end{abstract}

\section{Introduction}


Reticular materials are a class of crystalline, porous materials made of molecular building blocks that are linked by strong chemical bonds~\cite{yaghi2003reticular}.
They exhibit exceptional properties due to their highly porous structure, high surface area, tunable pore sizes and morphologies~\cite{lyu2020digital}.
Their versatility is evidenced by a broad range of industrial applications, among them heterogeneous catalysis~\cite{bavykina2020metal}, energy storage~\cite{zhao2018metal}, water treatment~\cite{li2022metal}, chemical sensing~\cite{kreno2012metal}, heat transfer~\cite{islamov2023high}, gas capture~\cite{moghadam2024progress} and drug delivery~\cite{horcajada2010porous}.

Following recent advances in generative AI, several models have been proposed to explore the large chemical space covered by reticular materials~\cite{yao2021inverse, park2024inverse, park2024generative, kang2024chatmof, cipcigan2024discovery}.
These models aim to generate reticular structures with optimized properties. Such structures are hypothetical as they have not yet been synthesised and tested in the lab. Devising a synthesis protocol for computationally generated structures requires a subject matter expert (SME). This is, however a challenging task given the large number of possible structures. An AI model that correlates a computationally discovered material with a lab synthesis protocol is, therefore, highly desirable. A first step towards the creation of such a model is building a database of existing synthesis protocols.


One approach for creating such database is applying information extraction techniques to the existing body of scientific literature. A large number of reticular materials have been reported in the literature alongside their respective synthesis protocols~\cite{moghadam2017development, moghadam2020targeted}. It is worth noting, however, that overlapping discoveries are common, given that the same material can be produced by means of different synthesis protocols~\cite{chung2019advances}. Transfer learning has been suggested as means to improve information extraction on existing corpora of scientific texts related to materials~\cite{data2020olivetti}. For example, fine-tuning techniques allow for adapting existing general-purpose AI models to specific tasks in domains for which comparatively little data exists. However, recent developments in LLMs have enabled information extraction based on prompt engineering and few-shot learning tasks~\cite{polak2024}.


In this paper, we propose using large language models (LLMs), without the need for additional training or fine-tuning, for extracting synthesis protocols of reticular materials from scientific literature, i.e., unstructured PDF documents. We use prompt engineering with in-context learning (ICL)~\cite{ICL_NEURIPS2023_73950f0e} for providing in the prompt all the context needed by the LLM to process the instructions. Together with instructions and input data, we provide examples that guide the LLM output production. This technique reduces the risk of hallucination, since all the context needed to execute the instruction is provided within the prompt. Also, it accelerates the process of information extraction because it does not require SME-based annotation of thousands of sentences/paragraphs for fine-tuning the models.

Our domain-independent Knowledge Extraction Pipeline (KEP) uses LLMs for extracting relevant information from PDF documents. The pipeline is composed of four main modules: (i) \emph{PDF extractor}: processes the PDF to extract the text; (ii) \emph{Paragraph classification}: processes the text in order to select only the relevant paragraphs (i.e., paragraphs that have the information the user is interested in); (iii) \emph{Information extraction}: processes the relevant paragraphs and extract the relevant information; and (iv) \emph{Knowledge representation}: interprets and assigns meaning to the information while representing the related knowledge. The pipeline uses LLMs with prompt-engineering and ICL in two modules, namely \emph{paragraph classification} and \emph{information extraction}, which are the focus of this paper. In addition, for identifying the best set of examples to be used in the prompts of these two modules, we propose the \emph{Examples selection} phase. This phase measures the performance of the LLMs in a given task and, by using different sets of examples, identifies the set to be used for optimal LLM performance.

We have used five families of LLMs in both \emph{paragraph classification} and \emph{information extraction} modules and have compared their performance. We note that these open-source LLMs are not domain-specific and were not fine-tuned for our tasks. Our experiments indicate that: (i) even without fine-tuning or training, some of these models have achieved high performance in case ICL was used to provide examples in the prompt; (ii) the examples used in the prompt affect model performance and, hence, must be chosen carefully; and (iii) the same set of examples may lead to varying results if used in different models.

Some recent papers share our work's objectives, however, they differ methodologically~\cite{polak2024,Huo2019,Kononova2019,park2022mining}. For example, Polak \textit{et al.} (2024) ~\cite{polak2024} reported a pipeline for extracting information from unstructured text in the material discovery domain using language models. However, the cited work focused on simple extraction tasks, e.g., \emph{material, value and unit}, while our pipeline is aimed at complex information associated with synthesis protocols that require additional classification. Unlike in our approach which is based on few-shot prompts providing examples for facilitating the information extraction, the cited work applies zero-shot methods for determining the relevance of sentences or paragraphs. Huo \textit{et al.} (2019)~\cite{Huo2019} introduced a semi-supervised machine learning approach for classifying inorganic materials synthesis steps in scientific papers. The authors used the Latent Dirichlet Allocation (LDA) unsupervised topic modeling algorithm for clustering terms that are typically used in synthesis descriptions. A random forest classifier, based on annotations of hundreds of paragraphs, categorized the occurring synthesis types. This approach also used a Markov chain for modeling the sequence of steps, creating flowcharts of synthesis procedures.  

In Kononova \textit{et al.} (2019)~\cite{Kononova2019}, the authors generated a dataset with “codified recipes” for solid-state synthesis which was automatically extracted from scientific publications using traditional text mining and natural language processing approaches. The authors used the two-step paragraph classification approach described in Huo \textit{et al.} (2019)~\cite{Huo2019} for finding paragraphs on solid-state synthesis. The extraction pipeline consisted of several algorithms (BiLSTM-CRF, Material Parser, etc.) for identifying materials related information, including synthesis steps and conditions. Compared to our method, the cited work required considerable annotation effort and employed a less straightforward extraction pipeline. We note that our method relies primarily on the LLM capabilities for text understanding, without specialized tokenizers or entity recognizers. Finally, Park \textit{et al.} (2022)~\cite{park2022mining} created a four-step pipeline, with text extraction from XML/HTML or PDF files and classifying relevant paragraphs, performing named entity recognition and, a fully connected multi-layer with dropout as classifier.

Another promising, less related approach is using ``AI chatbot agents'' for assisting materials scientists in specific pipeline tasks. In reference \cite{zheng2023chatgpt_chemistry_assistant}, the authors used prompt engineering for guiding a ChatGPT-based bot to extract MOF synthesis information from various sources. The authors leveraged a bot-like interface for answering questions about synthesis procedures and chemical reactions. In reference  \cite{zheng2023chatgpt_research_group}, the authors leveraged multiple AI assistants, such as LLMs and specific ML algorithms, as lab assistants to support a human SME, enabling productivity levels similar to those of an entire research team. While the approach was not fully automated, it provided a proof-of-concept of how language models can be leveraged for accelerating materials discovery. 

The remainder of this paper is organized as follows. 
Section~\ref{use-case} introduces the use case, Section~\ref{kep} describes in details the pipeline applied to the use case and Section~\ref{results} presents our experiments. 
Section~\ref{conclusion} concludes and presents some future work.
\section{Use Case: Synthesis Protocols of Reticular Materials}
\label{use-case}


With the goal of extracting knowledge about the synthesis of reticular materials, i.e., MOFs, ZIFs, COFs and zeolites, we have searched the scientific literature by using Elsevier's API\footnote{\url{https://github.com/ElsevierDev/elsapy}} and downloaded full-text PDFs from the SCOPUS database.\footnote{\url{https://www.scopus.com}}. Our approach is based on extracting information from PDFs, and not XMLs, since not always a XML file will be available for a given document. Notice that our extraction pipeline (see Section~\ref{kep}) was not created to manipulate only documents available in Elsevier, where their XML files are also provided, but to process any PDF document (including those that are images). 

Our search employed the following keywords and wildcard terms to capture relevant references: `MOF', `metal organic framework', `metal-organic framework', `metal-organic-framework', `COF', `covalent organic', `covalent-organic', `ZIF', and `zeolit* imidazol*'. We further limited the search to articles published in journals within Chemistry, Chemical Engineering, Materials Science, Energy, Engineering, Environmental Science, Physics and Astronomy, and Biochemistry, Genetics, and Molecular Biology, retrieving 6,669 articles.

The results were then filtered, by using the filter provided in the Elsevier API, to include only open-access articles with DOI identifiers from the following publishers: Elsevier (10.1016), Wiley Blackwell (10.1002), The Royal Society of Chemistry (10.1039), American Chemical Society (10.1021), Springer-Verlag (10.1007), Nature Publishing Group (10.1038), and MDPI (10.3390).

To create a public dataset, we finally kept only articles under the \texttt{CC-BY-4.0} or \texttt{CC-BY-3.0} licenses, resulting in 2,032 \texttt{CC-BY-4.0} articles and 255 \texttt{CC-BY-3.0} articles. These CCBY license papers were selected by performing web-scrapping from the list of DOIs provided by the Elsevier API. Since we are considering only papers with CCBY 3.0 and 4.0 licenses, everyone can retrieve the PDFs.

After collecting the data, we randomly selected 305 articles in PDF format~\footnote{171 articles with at least one paragraph describing a synthesis protocol and 134 articles without any synthesis protocol description.}. We then extracted from these PDFs 188 paragraphs describing synthesis protocols, and 137 examples of paragraphs not describing synthesis protocols (a total of 325 paragraphs).
This curated set of paragraphs constitutes our golden collection of classified paragraphs. For details about how those paragraphs were extracted, see Section~\ref{kep}.

Subsequently, a team of eleven research scientists (composed of 2 SMEs) annotated each of the synthesis-related paragraphs on a case-by-case basis for extracting the following information:
(i) the description of the synthesis product; (ii) the equipment used as an energy source; (iii) the conditions under which the synthesis occurred (e.g., reaction time, reaction temperature, current density); and (iv) the reactants and solvents used, including their descriptions, quantities, and units of measurement.

Intentionally, some paragraphs were selected for annotation by multiple SMEs, leading to some inconsistencies. These inconsistencies were then used to refine the annotation guidelines. The data was reviewed on a case-by-case basis by SMEs using a custom-built graphical interface 
and compiled in a final set of 131 syntheses descriptions encoded in a JSON format, thereby creating our golden dataset of annotated synthesis information. Table~\ref{tab:golden-dataset} summarizes the data in our golden dataset.

\begin{table}[h!]
\centering
\caption{Overview of golden dataset}
\label{tab:golden-dataset}
\begin{tabular}{llr}
\hline
 & Synthesis & Not Synthesis \\
\hline
paragraphs classified & 188 & 137 \\
annotated paragraphs  & 131 & - \\
\hline
\end{tabular}
\end{table}


\section{Knowledge Extraction Pipeline (KEP)}
\label{kep}

\begin{figure}
\centering
\includegraphics[width=.9\textwidth]{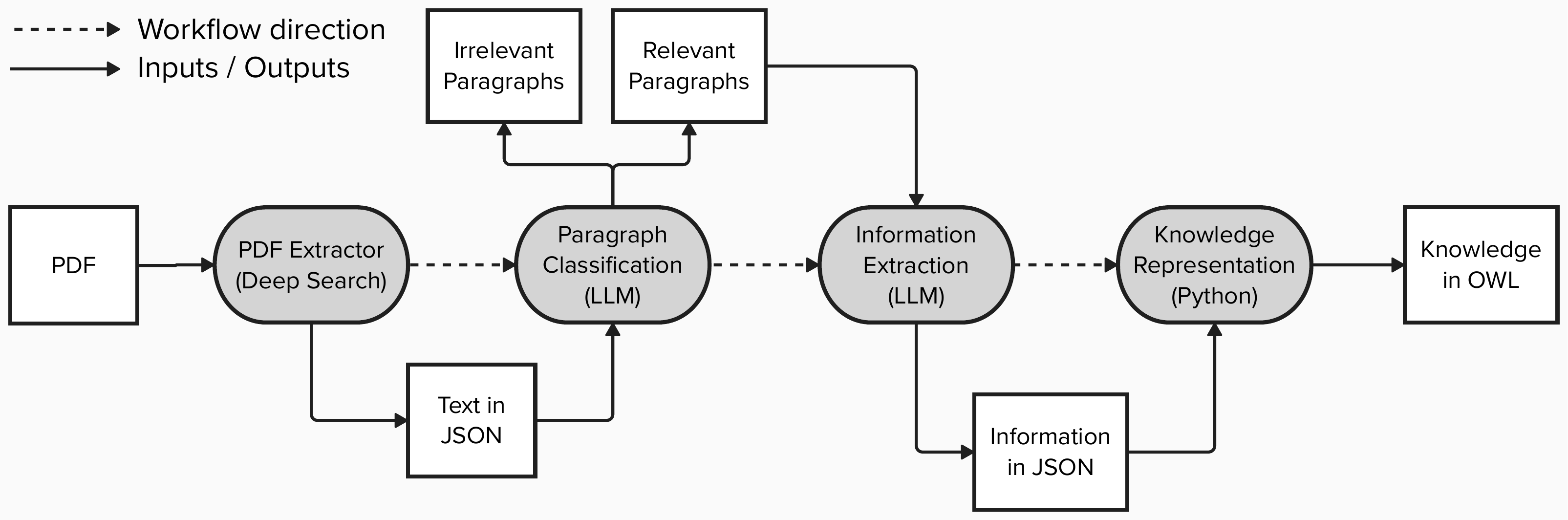}
\caption{Knowledge Extraction Pipeline (KEP) with the four KEP modules highlighted in gray color. Also shown are the respective inputs and outputs}\label{fig:kep_2}
\end{figure}

KEP is a domain-independent pipeline that helps extract knowledge from unstructured data. It is composed of four main modules: \emph{PDF extractor}, \emph{Paragraph classification}, \emph{Information extraction} and \emph{Knowledge representation}, as shown in Figure~\ref{fig:kep_2}. The \emph{PDF extractor} processes the PDF to extract paragraphs, since we assume that SMEs are interested in paragraphs containing specific information. The \emph{Paragraph classification} classifies the extracted paragraphs into \emph{relevant} or \emph{irrelevant}, according to the task the SME is interested in. When applying this module to our use case, \emph{relevant} paragraphs are those describing synthesis protocols of reticular materials. 

\emph{Information extraction} processes the relevant paragraphs and extracts the relevant information. When applying this module to our use case, the relevant information is the synthesis details such as the description of the synthesis product, the experimental conditions (such as reaction time and temperature), and the reagents and solvents used in the synthesis.
The final module, \emph{Knowledge representation}, interprets and assigns meaning to the extracted information while creates the knowledge representation. In the synthesis protocol use case, the knowledge representation is characterized by (i) the normalization of the unities; (ii) by the instantiation of entities of different kinds (such as productions, reactants and solvents), and (iii) by the instantiation of the relationships (such as used-reactant and used-solvent) that link the entities to the synthesis where they take part. For instance, it is possible to represent that the same reactant is being used in syntheses of two different products and that same product can be synthesized by two different synthesis.

The \emph{PDF extractor} was implemented using the DS4SD open-source tool\footnote{\url{https://ds4sd.github.io/}}
that converts unstructured PDF documents into JSON files containing the document elements such as section titles, paragraphs, footnotes,  headers, figure captions and tables, etc. DS4SD is also able to process PDFs that are indeed images since it uses an OCR engine to extract text-snippets from those images.
The \emph{Paragraph classification} and \emph{Information extraction} modules, which are the focus of this paper, were implemented by using 
open source LLMs of the Flan, Granite, LLaMa, Mistral and Mixtral families. 
As detailed in Section~\ref{results}, we compare the performance of these five families of LLMs when used in both the \emph{Paragraph classification} and \emph{Information extraction} modules. The LLMs were used without fine-tuning or training for the extraction of synthesis related information or on any task defined specifically for the Material Discovery domain. We only used prompt-engineering and ICL. 

To select the best set of examples to be provided in the prompt, the pipeline adds an additional step to each of the LLM's modules, namely \emph{Paragraph classification} and {Information extraction}. The \emph{Examples selection} step aims to select the best set of examples to be used in each tested LLMs for each one of the tasks, \emph{paragraph classification} and \emph{information extraction}, see Section~\ref{subsec:kep-examples_selection}.

\subsection{Paragraph classification}
\label{subsec:kep-paragraph-selection}

Since the goal of this module is the classification of paragraphs as \emph{relevant} or \emph{irrelevant}, the prompt to be used in this model should describe the difference between a relevant and an irrelevant paragraph. In addition, a sentence explicitly instructing the LLM that it should not provide an explanation together with the classification may be required. 

Since we are not using zero-shot prompting but ICL prompting, we not only provide the LLM with the aforementioned instructions, but also give it several examples of paragraphs and their corresponding classifications. In Section~\ref{results} we demonstrate that, by providing just a few examples in the prompt, the performance of the LLMs tends to increase significantly. Below is an example of instructions used, along with an example of paragraph\footnote{Paragraph extracted from~\cite{Matveevskaya2023}.} and its corresponding classification, also provided in the prompt. This paragraph was classified as "S" meaning it is a paragraph describing a synthesis protocol. 

\paragraph{Instruction:} \emph{You are assisting a chemist in classifying paragraphs from scientific articles. Mark the paragraph as 'S' if it describes the components of synthesis protocols for reticular materials, or 'I' if it does not include a synthesis description. After reviewing the examples, classify the given paragraph. Do not add any information or explanation besides 'S' or 'I' in the answer.}

\paragraph{Example:} \emph{"Synthesis of Zn-MOF:
Bis(imidazole-1-yl)methane was synthesized analogously to a the procedure reported in [43]. All other materials were obtained from commercial sources and were used as received.
\{[Zn(bim)(bdc)]0.8DMF0.4EtOH0.1H$_2$O\}$_n$ (Zn-MOF). Bis(imidazol-1-yl)methane (bim) (3.0 mg, 0.02 mmol), terephthalic acid (6.6 mg, 0.04 mmol), and Zn(NO3)2·6H2O (7.6 mg, 0.02 mmol) were dissolved in DMF/EtOH/H2O (2:1:1, vol.) mixture (1 mL), placed in a 4 mL screw-cap vial, and heated to 100 \textdegree C for 24 h."
\\
\\
Classification: S}


\subsection{Information extraction} 
\label{subsec:kep-information-extraction}

The prompt used in the \emph{Information extraction} module should inform to the LLM the kind of knowledge that should be extracted. In case of a complex structure, the prompt should suggest to the LLM to represent the extracted information following a given schema in well-known format, such as JSON~\cite{JSON}. It is reasonable to assume that the LLM will be able to parse this format since it is a commonly used data format that appeared in several documents used to train the LLM. In order to exemplify, find below the instruction we used and the JSON annotation related to the synthesis paragraph presented in Section~\ref{subsec:kep-paragraph-selection}. 

\paragraph{Instruction:} \emph{You are assisting a chemist in identifying and extracting descriptions of the synthesis of reticular materials from paragraphs. For each synthesis described in a paragraph, your task is to produce a JSON object that encodes the components involved in the synthesis, following the format provided in the examples. After reviewing the examples, carefully analyze the last paragraph and create a JSON object for each synthesis you find, ensuring that it adheres to the structure and conventions demonstrated.}

\paragraph{Example:} \emph{"Synthesis of Zn-MOF:
Bis(imidazole-1-yl)methane was synthesized analogously to a \dots screw-cap vial, and heated to 100 \textdegree C for 24 h."}
\begin{lstlisting}[frame=none, columns=fullflexible, numbers=none, backgroundcolor=\color{white}, escapeinside={(*}{*)}]
{"output": {
    "product": {
        "description": "Zn-MOF", 
        "material_type": "MOF", 
        "conditions": [
            {"description": "reaction temperature", "value": 100 , "unit": "oC"}, 
            {"description": "reaction time", "value": 24, "unit": "h"} 
        ]
    }, 
    "reactants": [
        {"description": "Bis(imidazol-1-yl)methane (bim)", "value": 0.02, "unit": "mmol"}, 
        {"description": "terephthalic acid", "value": 0.04, "unit": "mmol"},
        {"description": "Zn(NO3)2-6H2O", "value": 0.02, "unit": "mmol"} 
    ], 
    "solvents": [
        {"description": "DMF/EtOH/H2O (2:1:1, vol)", "value": 1.0, "unit": "mL"} 
    ]
}}
\end{lstlisting}

\subsection{Examples selection}
\label{subsec:kep-examples_selection}

It is well-known that the performance of LLMs to execute a given task is significantly influenced by the set of examples provided in the prompt. In addition, due to the different characteristics of how the LLMs were trained, it is expected that different LLMs will require different sets of examples to achieve their highest performance when executing the same task. 

Therefore, the \emph{Examples selection} step was included and associated with each KEP module that uses LLMs to help on the selection of the best set of prompt examples to be used. \emph{Examples selection} receives as input the model to be tested, a golden dataset and the number of examples to be selected as examples. It randomly selects from the dataset some instances to be used as examples in the prompt, and all other instances are used to measure the performance of the model. This step is executed for all possible combinations of examples or until the user is satisfied with the performance of the model in one of the executions. The set of examples that leads the LLM to achieve the highest performance is the one selected to be used in the associated KEP module.

\section{Experiments}
\label{results}

This section presents the experiments we ran with 5 families of open-source LLMs. None of them were trained or fine-tuned to extract synthesis details from paragraphs or to execute any specific task in the Material Discovery domain. We selected 2 models of each family\footnote{Exceptions: mistral and mixtral}
, prioritized the models that have been fine-tuned using a collection of instructions (not related to our tasks) and chosen the last released ones\footnote{Exception: llama-3-70b-instruct selected instead of llama-3-1-70b-instruct since it has a highest performance in the tasks we are testing.}. Ultimately, the selected models were: (i) flan: 
flan-t5-xxl-11b, flan-ul2-20b; (ii) granite: granite-20b-code-instruct, granite-34b-code-instruct; (iii) llama: llama-3-70b-instruct, llama-3.1-405b-instruct; (iv) mistral: mistral-large; and (v) mixtral: mixtral-8x7b-instruct-v01. See the description of each model in Appendix~\ref{app:models-description}.

\subsection{Examples selection}
\label{subsec:result-examples-selection}

\paragraph{Paragraph classification:}
From the original set of 325 classified paragraphs, we reduced the golden dataset by downselecting only 50 paragraphs to demonstrate that, even when testing the prompt examples selection in a small dataset, it is possible to achieve a good performance on a majority of the tested models. In addition, the use of a small dataset helps demonstrate that the approach does not require the manual classification/annotation of thousands or hundreds of examples. 

In the set of 50 paragraphs we ensure that 25 paragraphs are relevant (i.e., classified with "S" and mentioning synthesis protocol) and 25 are irrelevant (i.e., classified with "I" and not mentioning synthesis protocols). We fixed the number of examples to be provided in the prompt to 5, since paragraphs describing synthesis protocols are typically very large and the prompts have a limited number of tokens. Our goal is to find the best set of 5 examples used in the prompt that helps the models achieve their highest performance. The accuracy of each model was measured by using the F1 metric. 

For each model, we executed 100 runs by providing in the prompt the instruction mentioned in Section~\ref{subsec:kep-paragraph-selection} and 5 examples randomly selected from 50 possibilities. We tested the output with the remaining 45 paragraphs not provided in the prompt. Table~\ref{tab:examples-paragraph-selection-results} presents the result of our experiments. For each one of the models, the table indicates the number of paragraphs mentioning synthesis protocols ("S") and the number of irrelevant paragraphs ("I") used in both the worst and best prompt together with the F1 value for each case. 

\begin{table}[!ht]
\centering
\caption{The best-case (highlighted in bold) and worst-case (underlined) scenarios in the selection of examples to be used in the prompt of the \emph{Paragraph classification} module.}
\label{tab:examples-paragraph-selection-results}
\begin{tabular}{llllllr}
Model & Worst & & & Best & & \\
\hline
& \#S & \#I & F1 &\#S & \#I & F1 \\
\hline
flan-t5-xxl-11b  & 1 & 4 & 0.93 & 3 & 2 & \textbf{1.0} \\
flan-ul2-20b  & 3 & 2 & \underline{0.0} & 1 & 4 & 0.98 \\
granite-34b-code-instruct  & 1 & 4 & 0.30 & 2 & 3 & 0.92 \\
granite-20b-code-instruct  & 2 & 3 & 0.32 & 2 & 3 & 0.74 \\
llama-3-70b-instruct  & 1 & 4 & 0.71 & 4 & 1 & \textbf{1.0} \\
llama-3.1-405b-instruct  & 3 & 2 & \underline{0.0} & 3 & 2 & 0.95 \\
mistral-large  & 2 & 3 & 0.76 & 4 & 1 & \textbf{1.0} \\
mixtral-8x7b-instruct-v01  & 3 & 2 & 0.61 & 3 & 2 & \textbf{1.0} \\
\hline
\end{tabular}
\end{table}

The models with highest performance were flan-t5-xxl-11b, llama-3-70b-instruct, mistral-large and mixtral-8x7b-instruct-v01. Although llama-3-70b-instruct and mistral-large used the same number of relevant paragraphs and the same number of irrelevant paragraphs in their best cases, their prompts share only one paragraph (see Table~\ref{tab:json-paragraph-selection} in Appendix~\ref{app:examples-selection}). When testing the best prompt for mistral-large in llama-3-70b-instruct by using the same 45 testing examples, the performance of the model did not achieve F1=1.0, but F1=0.98. Although it is a small difference, it demonstrate that, different LLMs may need different examples in their prompts to achieve their highest performance. The models with worst performance were flan-ul2-20b and llama-3.1-405b-instruct. Although we included in the prompt a sentence stating that the answer should only include "S" or "I", their answers often also include an explanation; which we considered to be a hallucination and, thus, an incorrect answer.

\paragraph{Information extraction:}
The golden dataset used in this step is the 25 paragraphs mentioning synthesis protocols used in the previous step together with their coresponding JSON annotations. Different from the previous step, here we fixed the number of examples used in the prompt to 2, since the JSON annotation is being provided together with the paragraph, which significantly increases the number of tokens. Even with only 2 examples, flan-t5-xxl and flan-ul2 could not be tested since their prompt+result do not accept so many tokens~\footnote{Both flan models accept only 4,096 when comparing to llama that accepts 8,192}.

The experiment begun by randomly selecting 2 paragraphs+JSON to be used in the prompt for each one of the 6 models. For each model, we executed 100 runs by providing in the prompt the instructions mentioned in Section~\ref{subsec:kep-information-extraction} and the 2 examples of paragraph+JSON randomly selected from 25 possibilities. We tested the performance of the model with each prompt by using the 23 paragraphs that were not provided as examples in the prompt. The results are presented in Table~\ref{tab:examples-information-extraction-results}. To compare the JSON annotations provided by the LLM with the JSON annotations included in the golden dataset, a structure analysis based on each JSON key (i.e., name/value pair) was defined\footnote{To create a more fine-grained comparison between the JSONs, it would be necessary to compare their semantics and not only their structures, as different structures could have the same meaning.}.


\begin{table}[!ht]
\centering
\caption{The best-case and worst-case scenarios in the selection of examples of the \emph{Information extraction} module. The best results are highlighted in bold and the worst results are underlined.}
\label{tab:examples-information-extraction-results}
\begin{tabular}{llr}
\hline
Model  & Worst accuracy & Best accuracy \\
\hline
granite-34b-code-instruct  & 0.70 & 0.93 \\
granite-20b-code-instruct  & 0.65 & 0.84 \\
llama-3-70b-instruct  & 0.54 & \textbf{0.95} \\
llama-3.1-405b-instruct  & 0.53 & 0.94 \\
mistral-large & \underline{0.22} & 0.94 \\
mixtral-8x7b-instruct-v01 & 0.70 & 0.93 \\
\hline
\end{tabular}
\end{table}

The models that achieved the highest accuracy were llama-3-70b-instruct, llama-3.1-405b-instruct and mistral-large. However, it is important to notice that all of them achieved an accuracy higher than \textbf{0.84} even using only two examples in the prompt. Similar to what happened in the previous step, the experiments illustrate the influence of the examples in the accuracy of the model (E.g. llama-3.1-405b-instruct worst case was 0.53 and best case was 0.94). In addition, one of the paragraphs presented in the worst case of mistral-large appeared in the best case of mixtral-8x7b-instruct-v01 (see Table~\ref{tab:json-information-extraction} in Appendix~\ref{app:examples-selection}). Two related models that have the same example in opposite scenarios. Moreover, it is important to highlight that the two granites, the two llamas, and mixtral-8x7b-instruct-v01 included in their worst scenarios the same paragraph (see Table~\ref{tab:json-information-extraction} in Appendix~\ref{app:examples-selection}). It may indicate that there are examples that really do not help the models on executing their tasks.

\subsection{Paragraph classification}

After selecting the final set of five examples that maximize the performance of each model, the \emph{paragraph classification} module was tested by using the entire golden dataset of 275 paragraphs (325 minus the 50 used for prompt selection). For each model, the prompt was composed of the instructions mentioned in Section~\ref{subsec:kep-paragraph-selection} and the best set of examples selected for that model, as presented in Section~\ref{subsec:result-examples-selection}. Table~\ref{tab:paragraph-selection-results} summarizes the results for each model in terms of Precision, Recall, and F1 achieved with the best prompt. Llama-3-70b-instruct and mistral-large achieved \textbf{F1=0.98}. Although llama-3.1-405b-instruct and flan-ul2-20b have more parameters than the other model of their families, their performances were worse. It occurred due the hallucination mentioned in the \emph{Example section} step. Excluding granite-20b-code-instruct, all the models achieved \textbf{F1>0.84}, which is very good accuracy given that only five examples were provided in the prompt to these models. 

\begin{table}[!ht]
\centering
\caption{Experiments for the \emph{Paragraph classification} module (best results highlighted in bold).}
\label{tab:paragraph-selection-results}
\begin{tabular}{llllr}
Model & Precision & Recall & F1\\
\hline
flan-t5-xxl-11b & \textbf{0.98} & 0.96 & 0.97 \\
flan-ul2-20b & 0.96 & 0.96 &  0.96\\
granite-34b-code-instruct & 0.87 & 0.83 & 0.84 \\
granite-20b-code-instruct  & 0.75 & 0.70 & 0.72 \\
llama-3-70b-instruct & \textbf{0.98} & \textbf{0.98} & \textbf{0.98} \\ 
llama-3.1-405b-instruct & \textbf{0.98} & 0.83 & 0.88\\
mistral-large & \textbf{0.98} & \textbf{0.98} & \textbf{0.98}\\
mixtral-8x7b-instruct-v01 & 0.95 & 0.93 & 0.94 \\
\hline
\end{tabular}
\end{table}

\subsection{Information extraction}

This module was tested by using the golden dataset of 106 annotated paragraphs (131 minus the 25 used for prompt selection). For each model, the prompt was composed of the instructions mentioned in Section~\ref{subsec:kep-information-extraction} and the best set of examples selected for that model, as presented in Section~\ref{subsec:result-examples-selection}. Table~\ref{tab:information-extraction-results} summarizes the results of our experiments. The model that achieved the highest accuracy (\textbf{0.96}) was llama-3.1-405b-instruct, which is the biggest one. Other four models also achieved a very similar and high performance (mixtral-8x7b-instruct-v01, mistral-large, llama-3-70b-instruct and granite-34b-code-instruct). Notice that the smallest model (granite-20b-code-instruct) was the one that achieved the lower performance. The high accuracy achieved by the biggest models when compared to the smallest one is expected due to the complex of the task that involves the creation of a correct JSON. When considering both the \emph{Paragraph classification} and \emph{Information extraction} modules, the three models with highest performance and, thus, those that should be considered to be used in KEP to process all the selected papers mention in Section~\ref{use-case} are: llama-3-70b-instruct, mistral-large and mixtral-8x7b-instruct-v01.

\begin{table}[!ht]
\centering
\caption{Experiments for the \emph{Information extraction} module (best results highlighted in bold).}
\label{tab:information-extraction-results}
\begin{tabular}{llllr}
Model & Accuracy\\
\hline
granite-34b-code-instruct &0.93\\
granite-20b-code-instruct & 0.84\\
llama-3-70b-instruct &0.93\\
llama-3.1-405b-instruct &\textbf{0.96}\\
mistral-large & 0.95\\
mixtral-8x7b-instruct-v01 &0.94\\

\hline
\end{tabular}
\end{table}
\section{Conclusions and Future Research}
\label{conclusion}

In summary, we present a knowledge extraction pipeline for synthesis protocols of reticular materials that significantly reduces SME based classification and annotation tasks related to the training or fine-tuning of machine learning models. Our experimental results indicate that LLMs can achieve high performance with a limited set of examples within the prompt, even without training or fine-tuning the models for the specific domain. For example, by including five representative paragraphs in the prompt, we have reproducibly achieved F1=\textbf{0.98} in paragraph classification tasks. In information extraction tasks, we have used two paragraphs + JSON and llama-3.1-405b-instruct for achieving Accuracy=\textbf{0.96}. 

Our results highlight the necessity of testing different examples to be used in the prompt as this variation strongly influences model performance. For instance, in the \emph{Paragraph classification} module, the performance of mixtral-8x7b-instruct-v01, one of the best models in our study, ranges from F1=\textbf{0.61} to F1=\textbf{1.0}. In addition, the experiments show that different LLMs may require different sets of examples for achieving top performance. Although both llama-3-70b-instruct and mistral-large included four synthesis paragraphs and one irrelevant paragraph in their best set of examples, llama-3-70b-instruct has not achieved its highest performance with the best prompt chosen for mistral-large. Finally, a huge number of parameters in the model does not necessarily guarantee a superior model performance. Both flan-ul2 and llama-3.1-405b-instruct failed to achieve top performance in the classification of paragraphs if compared to other models of the same family.


Future research work should include comparative analyses with nonLLM methods in view of extraction time and quality, as well as measuring LLMs' performance for different materials applications. For creating a dataset of synthesis protocols for reticular materials, future research should address the following: (i) \textbf{refine JSONs comparison}: The creation of metrics for semantically comparing JSONs is needed to validate if the output of the model is structurally comparable with the golden dataset, and if it should be considered a valid JSON; (ii) \textbf{workflow extraction}: The extension of the \emph{Information extraction} module for extracting the synthesis workflow step-by-step; and (iii) \textbf{increase use case coverage}: The application of KEP to all paragraphs extracted from the selected 2,287 papers. Once processed, the resulting data set should be explored for analyzing the distributions of synthesis details made available in the scientific literature.

\bibliographystyle{IEEEtran}
\bibliography{references}

\begin{thebibliography}{10}
\providecommand{\url}[1]{#1}
\csname url@samestyle\endcsname
\providecommand{\newblock}{\relax}
\providecommand{\bibinfo}[2]{#2}
\providecommand{\BIBentrySTDinterwordspacing}{\spaceskip=0pt\relax}
\providecommand{\BIBentryALTinterwordstretchfactor}{4}
\providecommand{\BIBentryALTinterwordspacing}{\spaceskip=\fontdimen2\font plus
\BIBentryALTinterwordstretchfactor\fontdimen3\font minus
  \fontdimen4\font\relax}
\providecommand{\BIBforeignlanguage}[2]{{%
\expandafter\ifx\csname l@#1\endcsname\relax
\typeout{** WARNING: IEEEtran.bst: No hyphenation pattern has been}%
\typeout{** loaded for the language `#1'. Using the pattern for}%
\typeout{** the default language instead.}%
\else
\language=\csname l@#1\endcsname
\fi
#2}}
\providecommand{\BIBdecl}{\relax}
\BIBdecl

\bibitem{yaghi2003reticular}
O.~M. Yaghi, M.~O'Keeffe, N.~W. Ockwig, H.~K. Chae, M.~Eddaoudi, and J.~Kim,
  ``Reticular synthesis and the design of new materials,'' \emph{Nature}, vol.
  423, no. 6941, pp. 705--714, 2003.

\bibitem{lyu2020digital}
H.~Lyu, Z.~Ji, S.~Wuttke, and O.~M. Yaghi, ``Digital reticular chemistry,''
  \emph{Chem}, vol.~6, no.~9, pp. 2219--2241, 2020.

\bibitem{bavykina2020metal}
A.~Bavykina, N.~Kolobov, I.~S. Khan, J.~A. Bau, A.~Ramirez, and J.~Gascon,
  ``Metal--organic frameworks in heterogeneous catalysis: recent progress, new
  trends, and future perspectives,'' \emph{Chemical reviews}, vol. 120, no.~16,
  pp. 8468--8535, 2020.

\bibitem{zhao2018metal}
R.~Zhao, Z.~Liang, R.~Zou, and Q.~Xu, ``Metal-organic frameworks for
  batteries,'' \emph{Joule}, vol.~2, no.~11, pp. 2235--2259, 2018.

\bibitem{li2022metal}
R.~Li, N.~N. Adarsh, H.~Lu, and M.~Wriedt, ``Metal-organic frameworks as
  platforms for the removal of per-and polyfluoroalkyl substances from
  contaminated waters,'' \emph{Matter}, vol.~5, no.~10, pp. 3161--3193, 2022.

\bibitem{kreno2012metal}
L.~E. Kreno, K.~Leong, O.~K. Farha, M.~Allendorf, R.~P. Van~Duyne, and J.~T.
  Hupp, ``Metal--organic framework materials as chemical sensors,''
  \emph{Chemical reviews}, vol. 112, no.~2, pp. 1105--1125, 2012.

\bibitem{islamov2023high}
M.~Islamov, H.~Babaei, R.~Anderson, K.~B. Sezginel, J.~R. Long, A.~J.
  McGaughey, D.~A. Gomez-Gualdron, and C.~E. Wilmer, ``High-throughput
  screening of hypothetical metal-organic frameworks for thermal
  conductivity,'' \emph{npj Computational Materials}, vol.~9, no.~1, p.~11,
  2023.

\bibitem{moghadam2024progress}
P.~Z. Moghadam, Y.~G. Chung, and R.~Q. Snurr, ``Progress toward the
  computational discovery of new metal--organic framework adsorbents for energy
  applications,'' \emph{Nature Energy}, vol.~9, no.~2, pp. 121--133, 2024.

\bibitem{horcajada2010porous}
P.~Horcajada, T.~Chalati, C.~Serre, B.~Gillet, C.~Sebrie, T.~Baati, J.~F.
  Eubank, D.~Heurtaux, P.~Clayette, C.~Kreuz \emph{et~al.}, ``Porous
  metal--organic-framework nanoscale carriers as a potential platform for drug
  delivery and imaging,'' \emph{Nature materials}, vol.~9, no.~2, pp. 172--178,
  2010.

\bibitem{yao2021inverse}
Z.~Yao, B.~S{\'a}nchez-Lengeling, N.~S. Bobbitt, B.~J. Bucior, S.~G.~H. Kumar,
  S.~P. Collins, T.~Burns, T.~K. Woo, O.~K. Farha, R.~Q. Snurr \emph{et~al.},
  ``Inverse design of nanoporous crystalline reticular materials with deep
  generative models,'' \emph{Nature Machine Intelligence}, vol.~3, no.~1, pp.
  76--86, 2021.

\bibitem{park2024inverse}
H.~Park, S.~Majumdar, X.~Zhang, J.~Kim, and B.~Smit, ``Inverse design of
  metal--organic frameworks for direct air capture of co 2 via deep
  reinforcement learning,'' \emph{Digital Discovery}, vol.~3, no.~4, pp.
  728--741, 2024.

\bibitem{park2024generative}
H.~Park, X.~Yan, R.~Zhu, E.~A. Huerta, S.~Chaudhuri, D.~Cooper, I.~Foster, and
  E.~Tajkhorshid, ``A generative artificial intelligence framework based on a
  molecular diffusion model for the design of metal-organic frameworks for
  carbon capture,'' \emph{Communications Chemistry}, vol.~7, no.~1, p.~21,
  2024.

\bibitem{kang2024chatmof}
Y.~Kang and J.~Kim, ``Chatmof: an artificial intelligence system for predicting
  and generating metal-organic frameworks using large language models,''
  \emph{Nature Communications}, vol.~15, no.~1, p. 4705, 2024.

\bibitem{cipcigan2024discovery}
F.~Cipcigan, J.~Booth, R.~N.~B. Ferreira, C.~R. dos Santos, and M.~Steiner,
  ``Discovery of novel reticular materials for carbon dioxide capture using
  gflownets,'' \emph{Digital Discovery}, vol.~3, no.~3, pp. 449--455, 2024.

\bibitem{moghadam2017development}
P.~Z. Moghadam, A.~Li, S.~B. Wiggin, A.~Tao, A.~G. Maloney, P.~A. Wood, S.~C.
  Ward, and D.~Fairen-Jimenez, ``Development of a cambridge structural database
  subset: a collection of metal--organic frameworks for past, present, and
  future,'' \emph{Chemistry of Materials}, vol.~29, no.~7, pp. 2618--2625,
  2017.

\bibitem{moghadam2020targeted}
P.~Z. Moghadam, A.~Li, X.-W. Liu, R.~Bueno-Perez, S.-D. Wang, S.~B. Wiggin,
  P.~A. Wood, and D.~Fairen-Jimenez, ``Targeted classification of
  metal--organic frameworks in the cambridge structural database (csd),''
  \emph{Chemical science}, vol.~11, no.~32, pp. 8373--8387, 2020.

\bibitem{chung2019advances}
Y.~G. Chung, E.~Haldoupis, B.~J. Bucior, M.~Haranczyk, S.~Lee, H.~Zhang, K.~D.
  Vogiatzis, M.~Milisavljevic, S.~Ling, J.~S. Camp \emph{et~al.}, ``Advances,
  updates, and analytics for the computation-ready, experimental metal--organic
  framework database: Core mof 2019,'' \emph{Journal of Chemical \& Engineering
  Data}, vol.~64, no.~12, pp. 5985--5998, 2019.

\bibitem{data2020olivetti}
\BIBentryALTinterwordspacing
E.~A. Olivetti, J.~M. Cole, E.~Kim, O.~Kononova, G.~Ceder, T.~Y.-J. Han, and
  A.~M. Hiszpanski, ``{Data-driven materials research enabled by natural
  language processing and information extraction},'' \emph{Applied Physics
  Reviews}, vol.~7, no.~4, p. 041317, 12 2020. [Online]. Available:
  \url{https://doi.org/10.1063/5.0021106}
\BIBentrySTDinterwordspacing

\bibitem{polak2024}
\BIBentryALTinterwordspacing
M.~P. Polak and D.~Morgan, ``Extracting accurate materials data from research
  papers with conversational language models and prompt engineering,''
  \emph{Nature Communications}, vol.~15, no.~1, p. 1569, 2024. [Online].
  Available: \url{https://doi.org/10.1038/s41467-024-45914-8}
\BIBentrySTDinterwordspacing

\bibitem{ICL_NEURIPS2023_73950f0e}
\BIBentryALTinterwordspacing
N.~Wies, Y.~Levine, and A.~Shashua, ``The learnability of in-context
  learning,'' in \emph{Advances in Neural Information Processing Systems},
  A.~Oh, T.~Naumann, A.~Globerson, K.~Saenko, M.~Hardt, and S.~Levine, Eds.,
  vol.~36.\hskip 1em plus 0.5em minus 0.4em\relax Curran Associates, Inc.,
  2023, pp. 36\,637--36\,651. [Online]. Available:
  \url{https://proceedings.neurips.cc/paper_files/paper/2023/file/73950f0eb4ac0925dc71ba2406893320-Paper-Conference.pdf}
\BIBentrySTDinterwordspacing

\bibitem{Huo2019}
\BIBentryALTinterwordspacing
H.~Huo, Z.~Rong, O.~Kononova, W.~Sun, T.~Botari, T.~He, V.~Tshitoyan, and
  G.~Ceder, ``Semi-supervised machine-learning classification of materials
  synthesis procedures,'' \emph{npj Computational Materials}, vol.~5, no.~1,
  p.~62, Jul 2019. [Online]. Available:
  \url{https://doi.org/10.1038/s41524-019-0204-1}
\BIBentrySTDinterwordspacing

\bibitem{Kononova2019}
\BIBentryALTinterwordspacing
O.~Kononova, H.~Huo, T.~He, Z.~Rong, T.~Botari, W.~Sun, V.~Tshitoyan, and
  G.~Ceder, ``Text-mined dataset of inorganic materials synthesis recipes,''
  \emph{Scientific Data}, vol.~6, no.~1, p. 203, Oct 2019. [Online]. Available:
  \url{https://doi.org/10.1038/s41597-019-0224-1}
\BIBentrySTDinterwordspacing

\bibitem{park2022mining}
H.~Park, Y.~Kang, W.~Choe, and J.~Kim, ``Mining insights on metal--organic
  framework synthesis from scientific literature texts,'' \emph{Journal of
  Chemical Information and Modeling}, vol.~62, no.~5, pp. 1190--1198, 2022.

\bibitem{zheng2023chatgpt_chemistry_assistant}
Z.~Zheng, O.~Zhang, C.~Borgs, J.~T. Chayes, and O.~M. Yaghi, ``Chatgpt
  chemistry assistant for text mining and the prediction of mof synthesis,''
  \emph{Journal of the American Chemical Society}, vol. 145, no.~32, pp.
  18\,048--18\,062, 2023.

\bibitem{zheng2023chatgpt_research_group}
Z.~Zheng, O.~Zhang, H.~L. Nguyen, N.~Rampal, A.~H. Alawadhi, Z.~Rong,
  T.~Head-Gordon, C.~Borgs, J.~T. Chayes, and O.~M. Yaghi, ``Chatgpt research
  group for optimizing the crystallinity of mofs and cofs,'' \emph{ACS Central
  Science}, vol.~9, no.~11, pp. 2161--2170, 2023.

\bibitem{Matveevskaya2023}
\BIBentryALTinterwordspacing
V.~V. Matveevskaya, D.~I. Pavlov, A.~A. Ryadun, V.~P. Fedin, and A.~S. Potapov,
  ``Synthesis, crystal structure, and luminescent sensing properties of a
  supramolecular 3d zinc(ii) metal–organic framework with terephthalate and
  bis(imidazol-1-yl)methane linkers,'' \emph{Inorganics}, vol.~11, no.~7, p.
  264, Jun. 2023. [Online]. Available:
  \url{http://dx.doi.org/10.3390/inorganics11070264}
\BIBentrySTDinterwordspacing

\bibitem{JSON}
\BIBentryALTinterwordspacing
\emph{{ECMA-404}: The JSON data interchange syntax}, {ECMA International} Std.
  404, 2017. [Online]. Available:
  \url{https://ecma-international.org/publications-and-standards/standards/ecma-404/}
\BIBentrySTDinterwordspacing

\bibitem{flan-T5}
H.~W. Chung, L.~Hou, S.~Longpre, B.~Zoph, Y.~Tay, W.~Fedus, E.~Li, X.~Wang,
  M.~Dehghani, S.~Brahma \emph{et~al.}, ``Scaling instruction-finetuned
  language models,'' \emph{arXiv preprint arXiv:2210.11416}, 2022.

\bibitem{flan-UL2}
Y.~Tay, H.~W. Chung, L.~Hou, B.~Zoph, S.~Borgeaud, P.~He, S.~Narang, W.~Fedus,
  and D.~G. Patil, ``Ul2 20b: An open-source unified language learner model,''
  \emph{arXiv preprint arXiv:2301.07520}, 2023.

\bibitem{granite2024code}
\BIBentryALTinterwordspacing
I.~Research, ``Granite code models: A family of open foundation models for code
  intelligence,'' \emph{IBM Documentation}, 2024. [Online]. Available:
  \url{https://www.ibm.com/granite/playground/code/}
\BIBentrySTDinterwordspacing

\bibitem{llama-3-70B-Instruct}
\BIBentryALTinterwordspacing
\emph{Llama-3-70B-Instruct}, Meta Std., 2024. [Online]. Available:
  \url{https://huggingface.co/meta-llama/Meta-Llama-3-70B-Instruct}
\BIBentrySTDinterwordspacing

\bibitem{llama-3.1-405B}
\BIBentryALTinterwordspacing
\emph{Llama-3.1-405B}, Meta Std., 2024. [Online]. Available:
  \url{https://huggingface.co/meta-llama/Llama-3.1-405B}
\BIBentrySTDinterwordspacing

\bibitem{mistral2023comparative}
\BIBentryALTinterwordspacing
H.-C. Tsai, Y.-F. Huang, and C.-W. Kuo, ``Comparative analysis of automatic
  literature review using mistral large language model and human reviewers,''
  \emph{Sciety}, 2024. [Online]. Available:
  \url{https://sciety.org/articles/activity/10.21203/rs.3.rs-4022248/v1}
\BIBentrySTDinterwordspacing

\bibitem{mistral2024smixtral}
\BIBentryALTinterwordspacing
M.~AI, ``Sparse mixture of experts in large language models: Mixtral 8x7b,''
  \emph{arXiv preprint arXiv:2401.04088}, 2024. [Online]. Available:
  \url{https://arxiv.org/abs/2401.04088}
\BIBentrySTDinterwordspacing

\end{thebibliography}

\appendix

\section{Models Description}
\label{app:models-description}

Flan-T5~\cite{flan-T5}is a variant of the T5 (Text-to-Text Transfer Transformer) model, further fine-tuned using a mixture of instruction-based learning tasks. Like the original T5, Flan-T5 leverages a transformer architecture, specifically designed for text-to-text tasks, which means it treats both the input and output as text sequences, regardless of the task (e.g., translation, summarization, question-answering). The “Flan” component (Fine-tuned LAnguage Net) introduces instruction tuning, where the model is exposed to a variety of natural language instructions during its fine-tuning phase. This method allows the model to generalize better across tasks by learning to follow explicit human instructions. In essence, Flan-T5 adapts the standard pre-training and fine-tuning methods of T5 but adds an additional layer of task diversity through its instruction-based training. This approach enhances its performance on zero-shot and few-shot learning tasks, making it more versatile for a wide range of NLP applications.

Flan-UL2 (Unified Language Learner)~\cite{flan-UL2} is a variant of the UL2 architecture, designed for improved instruction-based fine-tuning similar to Flan-T5. UL2 is an advanced architecture that introduces a novel pre-training method utilizing a mixture of denoising tasks with different difficulty levels. This approach allows the model to adapt to a wider range of NLP tasks by balancing between simple and complex learning objectives. In the case of Flan-UL2, this model takes UL2 and further enhances it with instruction tuning, similar to the Flan-T5 approach. It is trained on a large variety of instruction tasks, making it highly proficient at zero-shot and few-shot learning across many tasks, such as summarization, translation, and question answering. The model’s ability to generalize across these tasks is further improved by the fine-tuning process with diverse datasets of instructions, allowing it to better understand human-like queries and execute complex tasks. This makes Flan-UL2 particularly powerful for applications requiring high versatility and adaptability in natural language understanding.

Granite-20B-Code-Instruct and Granite-34B-Code-Instruct~\cite{granite2024code} are part of the Granite family of large language models (LLMs) designed specifically for code-related tasks. Both models are fine-tuned versions of their respective base models, Granite-20B-Code-Base and Granite-34B-Code-Base, using instruction-based datasets to improve their ability to follow natural language instructions. These models, developed by IBM Research, are built for tasks such as code generation, bug fixing, code explanation, and translation across a wide range of programming languages, making them versatile tools for code-centric applications.
Granite-20B-Code-Instruct, with 20 billion parameters, was trained on trillions of tokens from various sources, including high-quality code, mathematical data, and instructional prompts. Its fine-tuning emphasizes logical reasoning and problem-solving, with a focus on generating and explaining code, alongside supporting tasks like API calling and debugging .
Granite-34B-Code-Instruct, with 34 billion parameters, extends these capabilities by being a more computationally powerful model, trained on a larger and more diverse dataset of code instructions. It can handle more complex coding tasks and demonstrates state-of-the-art performance across benchmarks for code synthesis, explanation, and debugging .
Both models are decoder-only architectures, optimized for generating human-readable code outputs from natural language inputs, and are trained with instruction tuning to improve their accuracy in code-based applications.

Llama-3-70B-Instruct~\cite{llama-3-70B-Instruct} is part of Meta’s Llama 3 family of large language models, specifically designed for instruction-following tasks. The model contains 70 billion parameters and is optimized for generating text in response to user prompts. It is a decoder-only model, which uses an optimized transformer architecture. The instruction-tuned version of Llama-3-70B benefits from Supervised Fine-Tuning (SFT) and Reinforcement Learning with Human Feedback (RLHF) to align its outputs with human preferences for helpfulness and safety. This fine-tuning process makes it particularly suitable for assistant-like applications, such as chatbots and task-oriented dialogue systems.
Llama-3-70B-Instruct was trained on an extensive corpus of 15 trillion tokens from publicly available datasets and supports a wide range of use cases, including multilingual text generation and code-related tasks. It incorporates improvements like Grouped-Query Attention (GQA) for faster inference and an expanded 8,192 token context window, allowing it to handle longer inputs effectively. The model has been tested extensively for safety, and Meta has integrated safeguards to limit misuse, including rigorous red teaming and cybersecurity assessments.
The model is available under the Meta Llama 3 Community License for both commercial and research applications. It’s praised for outperforming other models in several benchmarks, demonstrating significant advancements in multilingual dialogue capabilities and code generation.

Llama 3.1-405B-Instruct~\cite{llama-3.1-405B} is the largest model in the Llama 3.1 series by Meta, designed to provide state-of-the-art performance in multilingual dialogue and complex instruction-following tasks. With 405 billion parameters, it utilizes a transformer-based, decoder-only architecture optimized for extensive text generation tasks. It introduces enhancements in context handling, supporting up to 128,000 tokens, which makes it ideal for tasks like document summarization and long-context conversation .
This model is fine-tuned using a combination of Supervised Fine-Tuning (SFT) and Reinforcement Learning with Human Feedback (RLHF), enabling it to align better with human preferences and improve the safety and helpfulness of its outputs . Llama 3.1-405B was trained on a mixture of publicly available datasets containing approximately 15 trillion tokens, and its fine-tuning included more than 25 million synthetically generated instruction-based examples . Furthermore, it offers improved multilingual support beyond English, covering languages like German, French, Italian, Portuguese, Hindi, Spanish, and Thai .
The model is open-source and available under Meta’s custom open model license, encouraging use in both research and commercial applications .

Mistral AI’s large language models, particularly Mistral Large 2~\cite{mistral2023comparative}, represent significant advancements in both computational efficiency and reasoning capabilities. This model, featuring 123 billion parameters, is designed for tasks that require extensive reasoning, such as multilingual text processing, code generation, and mathematical problem-solving. With support for over 80 coding languages and a context window of 128,000 tokens, it excels in handling large documents and long, complex inputs.
Mistral Large 2 is particularly strong in benchmarks like MMLU (Massive Multitask Language Understanding), where it achieves an accuracy of 84

Mixtral-8x7B-Instruct-v0.1~\cite{mistral2024smixtral} is an advanced sparse mixture-of-experts (SMoE) model developed by Mistral AI. It incorporates a unique architecture where each layer contains eight experts (feedforward blocks), but only two are activated for each token during inference. This selective processing allows the model to manage a large number of parameters—47 billion in total—while only using 13 billion active parameters per token, which significantly reduces computation costs during inference.
Mixtral-8x7B-Instruct has been fine-tuned for instruction-following tasks through a combination of supervised fine-tuning (SFT) and Direct Preference Optimization (DPO). This model excels in benchmarks such as MMLU and GSM8K, matching or outperforming larger models like GPT-3.5 Turbo and Llama 2 70B in several areas, particularly code generation, reasoning, and multilingual tasks. Its ability to handle long sequences with a 32k token context window makes it highly effective for long-range information retrieval and complex prompts.

\section{Examples selection}
\label{app:examples-selection}
\paragraph{Paragraph classification} Table~\ref{tab:json-paragraph-selection} shows excerpts of JSONs with best and worst paragraphs selected as examples for each model. It is possible to see that few paragraphs appear in more than one prompt.

\newlength{\jsonfigWidth}  
\setlength{\jsonfigWidth}{.5\textwidth}  

\begin{longtable}{ | c | c | }
\caption{Partial JSONs with the best and worse examples for \emph{Paragraph classification} module.  \label{tab:json-paragraph-selection}}\\
\hline
\endfirsthead
\caption[]{(continued)}\\
\endhead
\includegraphics[width=\jsonfigWidth]{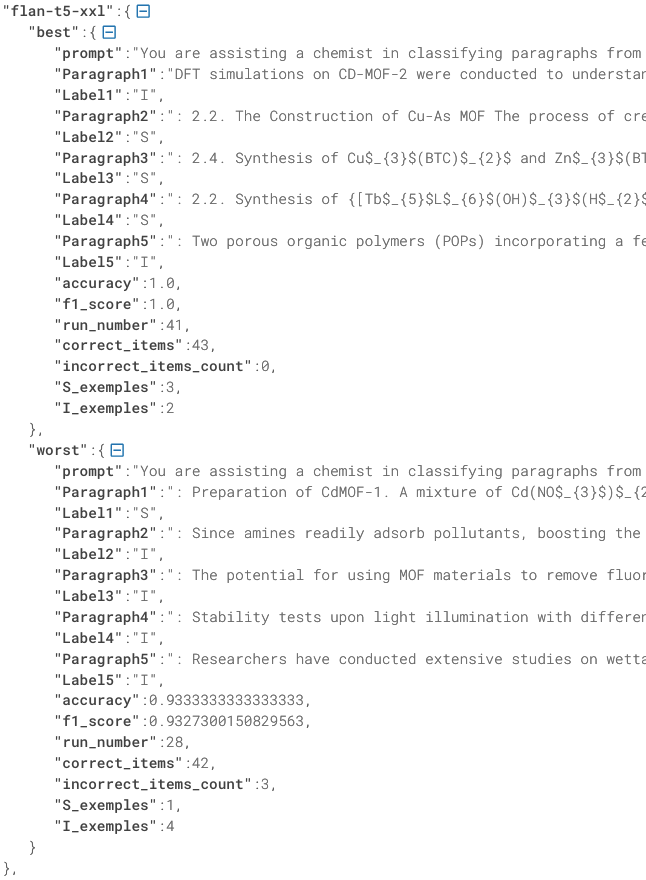} & 
\includegraphics[width=\jsonfigWidth]{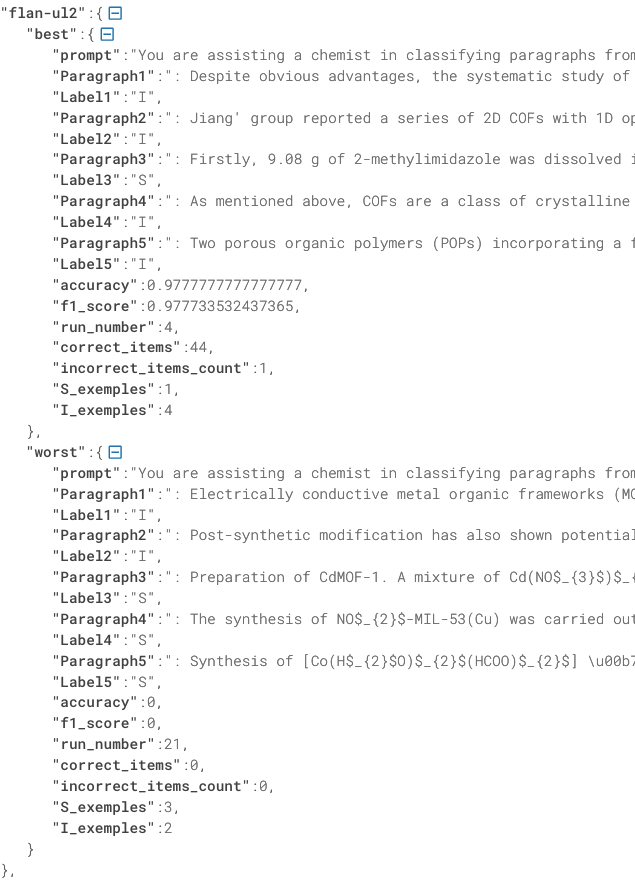}\\
\hline
\includegraphics[width=\jsonfigWidth]{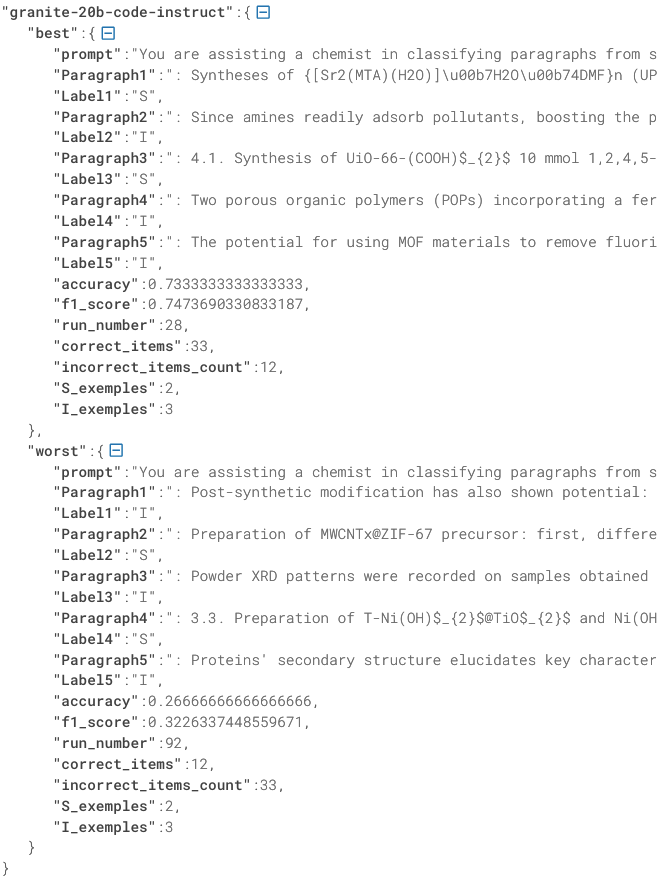} & 
\includegraphics[width=\jsonfigWidth]{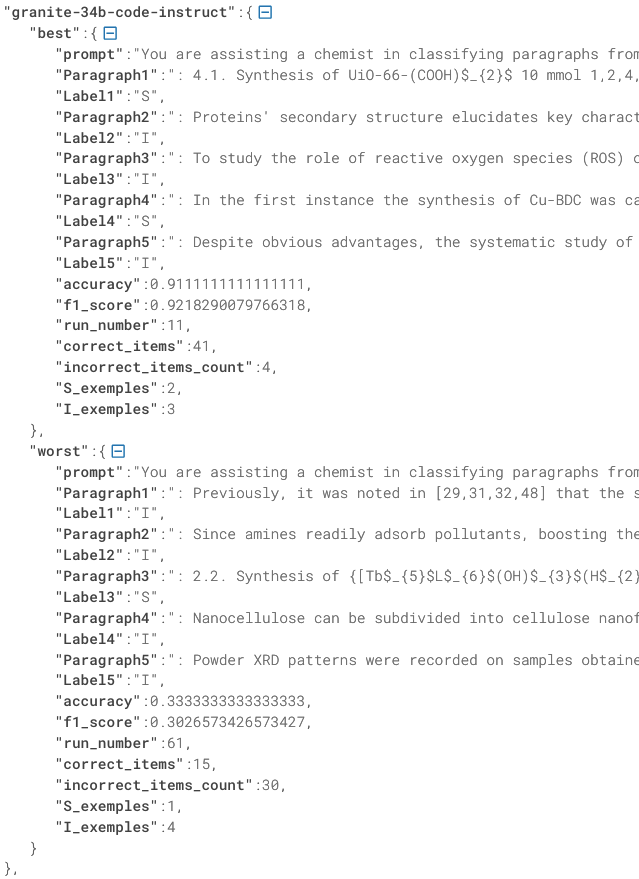}\\
\hline
\includegraphics[width=\jsonfigWidth]{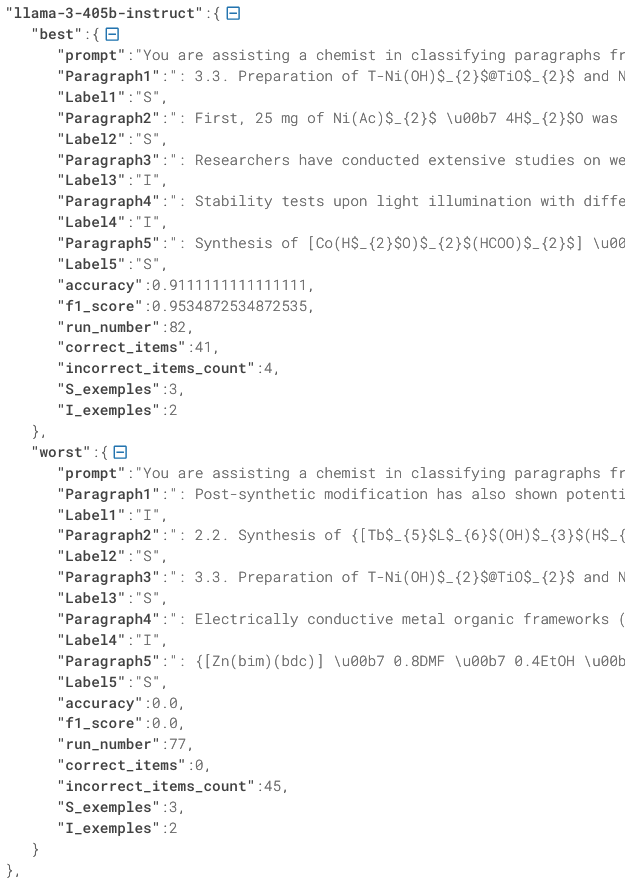} &
\includegraphics[width=\jsonfigWidth]{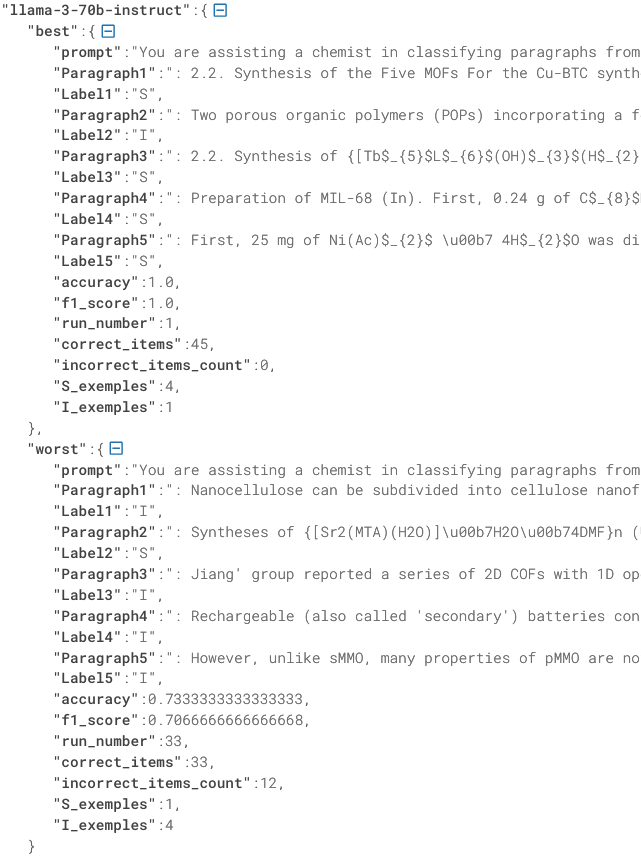} \\
\hline
\includegraphics[width=\jsonfigWidth]{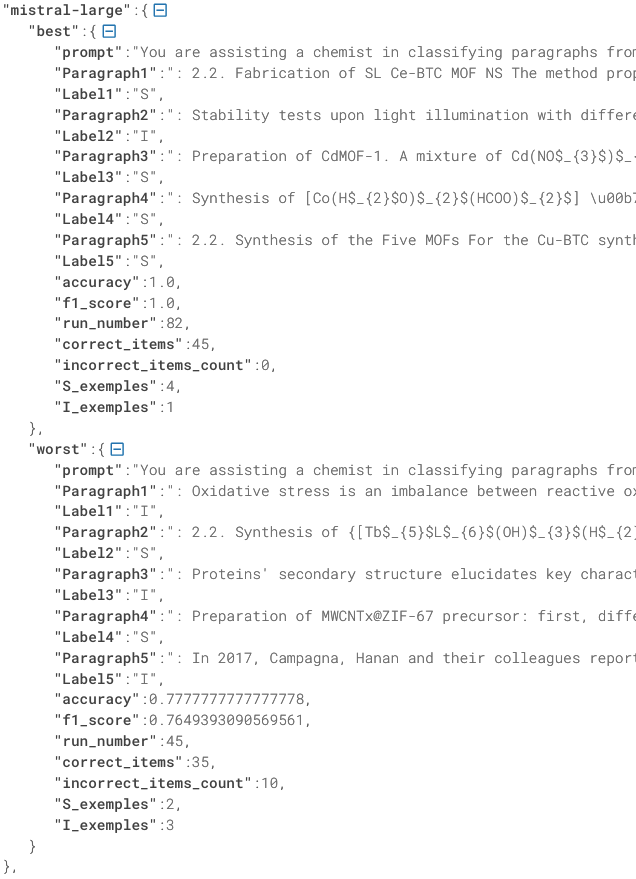} &
\includegraphics[width=\jsonfigWidth]{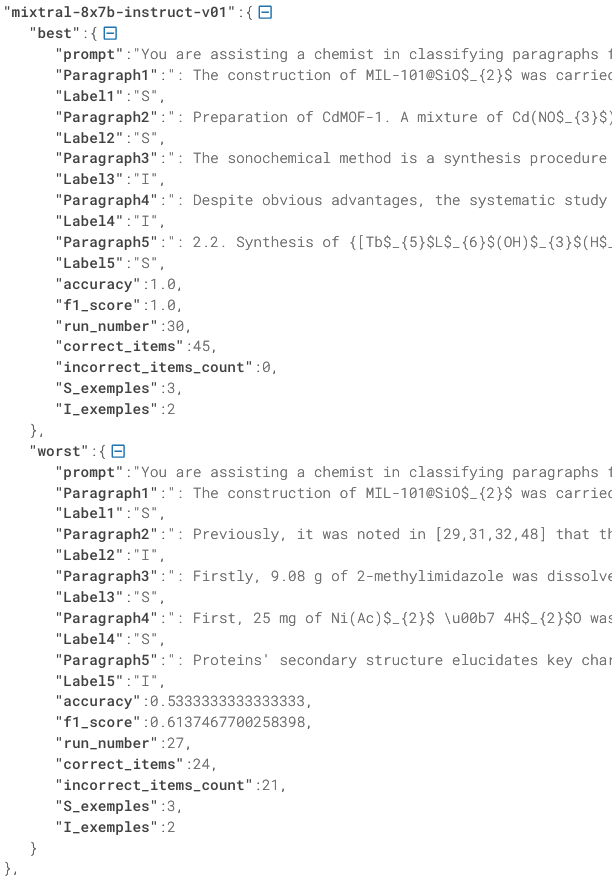} \\
\hline
\end{longtable}

\paragraph{Information extraction} Table~\ref{tab:json-information-extraction} shows the JSONs that include the best and worst paragraphs selected as examples for each model.

\setlength{\jsonfigWidth}{.8\textwidth}  

\begin{longtable}{ | c | }
\caption{Partial JSON with the best and worse examples for \emph{Information extraction} module.\label{tab:json-information-extraction}} \\
\hline
\endfirsthead
\caption[]{(continued)}\\
\endhead
\includegraphics[width=\jsonfigWidth]{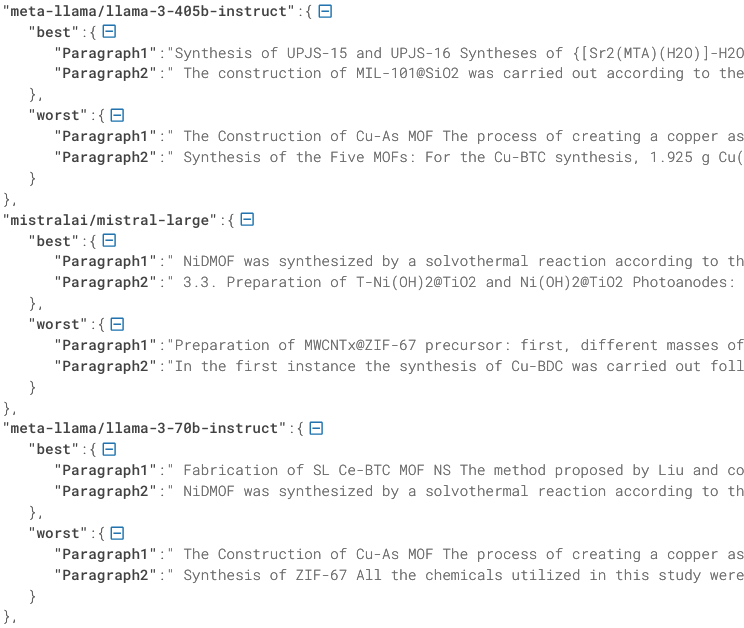}\\ 
\includegraphics[width=\jsonfigWidth]{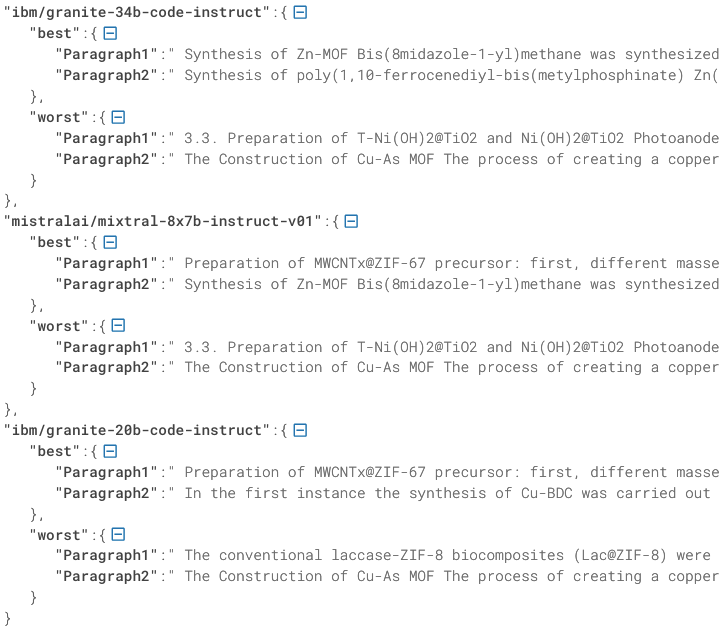}\\
\hline
\end{longtable}

\end{document}